\documentclass[11pt,a4paper]{article}
\usepackage{jheppub}
\usepackage{amssymb}
\usepackage{amsmath}
\usepackage{graphicx}
\usepackage{tikz}

\newcommand{\dd}{\mathrm{d}}

\title{Perturbative PDF of the total magnetization of the 4D Ising model }

\author[a]{Andrea Allais}
\affiliation[a]{Independent scholar}
\emailAdd{a\_allais@alum.mit.edu}

\abstract{We compute, at one loop in perturbation theory, the probability
density function of the total magnetization $M$ of the Ising model on the
4-torus and the 4-sphere. We develop a single perturbative expansion that is
valid in the symmetric phase as well as the broken symmetry phase, provided that the
correlation length is large compared to the system size $L$. We find that, at
the critical point, for large system size in lattice units, the PDF approaches
$p(M)\sim \exp(-f(L) M^4)$. Consequently, the critical value of the Binder
cumulant of the total magnetization is $U = 1 -
\frac{4\,\Gamma(5/4)^2}{3\,\Gamma(3/4)^2}$.  We validate our results by
comparison with Monte Carlo simulation.}
\begin{document}

\maketitle

\section{Main result}

When the system size $L$ is large, the probability distribution of the total
magnetization $M$ of the Ising model changes qualitatively between the two
phases of the model. In the symmetric phase, if the correlation
length $\xi$ is large in lattice units, but small compared to the system size,
the distribution is well approximated by a zero-mean normal.  In the broken
symmetry phase, still for $1 \ll \xi \ll L$, the distribution is well
approximated by a mixture of two normals centered at non-zero values $\pm
M_0$.

In the opposite regime, $1 \ll L \ll \xi$, in two and three dimensions, the
probability distribution of $M$ is a non-trivial function of $M$.  It is not
immediately clear if the same is also true in four dimensions. Since the field
theory that describes the Ising critical point in four dimensions becomes
weakly coupled at low energy, one may expect the probability distribution of
$M$ to remain a zero-centered normal even in this regime. This is in fact the
case for the total magnetization of a subsystem of intermediate size: large in
lattice units, but small compared to $L$ \cite{Binder1981}. Here we show that
the same is not true for the total magnetization of the entire system. We
compute the logarithm of the PDF of $M$ at one-loop in $4$ dimensions, and show
that, for $1 \ll L \ll \xi$, the distribution is not normal, but rather of the
form $-\log p(M)\sim f(L) M^4$.  We find that corrections to this form vanish
very slowly with increasing system size, like $1/\log L$.

\begin{figure}
\begin{center}
\includegraphics[scale=0.75]{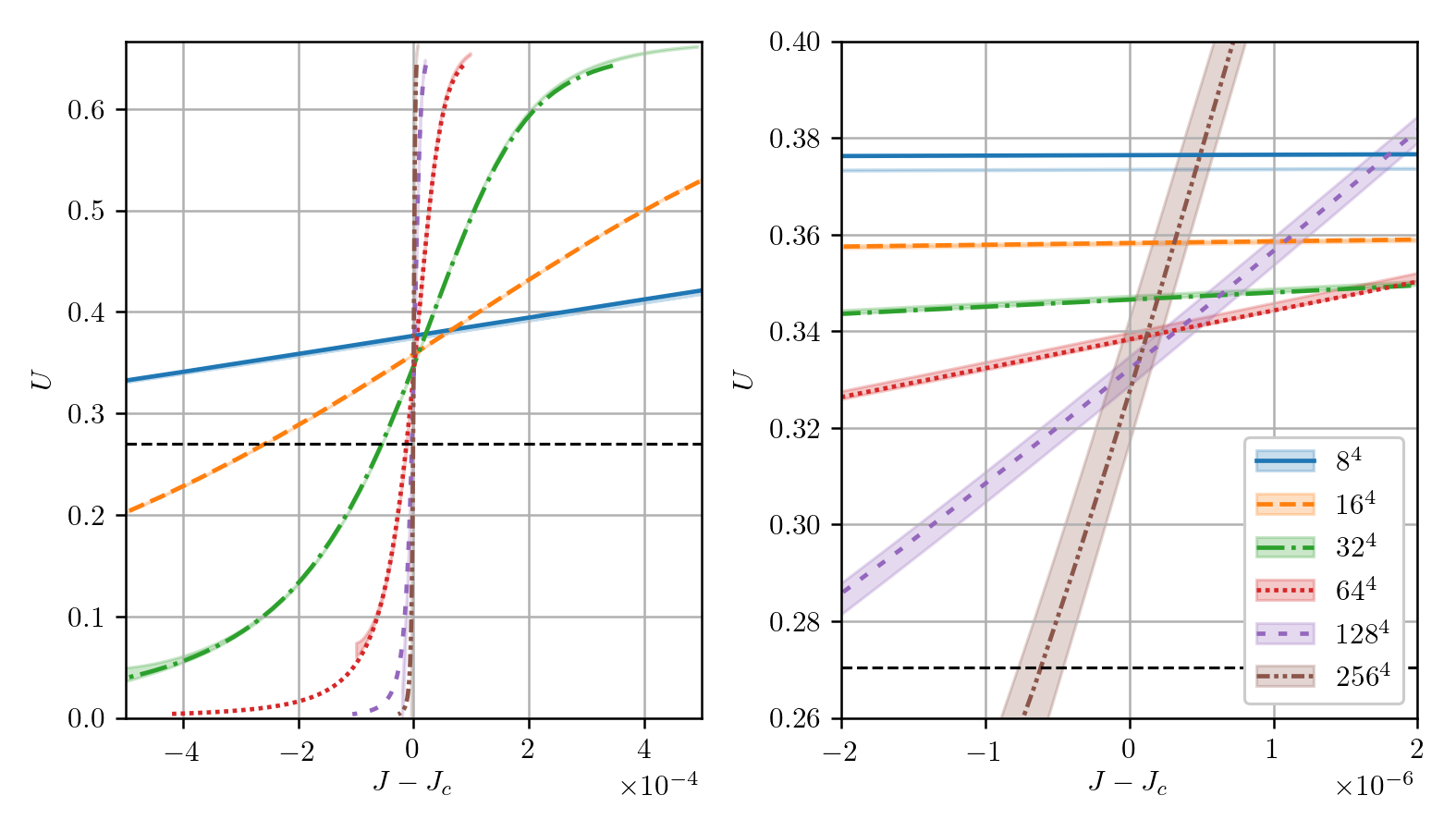}
\end{center}
    \caption{\label{fig:binder_cumulant} Binder cumulant of the total
    magnetization of the 4D Ising model. The left plot gives a broad picture,
    the right plot shows the critical region close-up.  The shaded regions
    display the one-sigma confidence intervals obtained from Monte Carlo
    simulation. The center lines are obtained from a 3-parameter fit of the
    perturbative result (\ref{eq:log_pdf}). The horizontal dashed line
    indicates the critical value (\ref{eq:critical_binder_cumulant}). The fit
    has $J_c = 0.1496938$, $J - J_c = -0.027\cdot r$, $g = 0.43$, with $r$, $g$
    defined at the renormalization scale $L = 2\pi R = 8$. }
\end{figure}

When the correlation length is large in lattice unit, the Ising model in 4
dimensions is described by the $\phi^4$ field theory:
\begin{equation}
    S[\phi] \equiv \int_{\mathcal{M}} \left(
    \frac{1}{2} \phi\left(\Delta + r_0\right)\phi +
    \frac{1}{24} u_0 \phi^4\right)\,.
\end{equation}

Here we take the manifold $\mathcal{M}$ to be a 4-torus of radius $R$,
\textit{i.e.} $x_i\sim x_i + 2\pi R$. In this field theory context, we define
the PDF of the average magnetization $m = M / V$ as:
\begin{equation}
    \label{eq:pdf_definition}
    p(m) \equiv \frac{1}{Z} \int \left[\mathcal{D}\phi\right]
    \delta\left(m - \frac{1}{V}\int_{\mathcal{M}} \phi\right)
    e^{-S[\phi]}\,,
\end{equation}
and we evaluate it at one loop in perturbation theory, obtaining:
\begin{equation}
\label{eq:log_pdf}
\begin{split}
    -\log p(m) = \mathrm{const.} +  V\Bigg(
  &\frac{1}{2} \frac{m^2}{R^2} \left(
    r R^2 + \frac{g}{6} \left(f_1\left(r R^2\right) - 1\right)
    + O\left(g^2\right)\right) + \\
  & \frac{2\pi^2}{9}gm^4\left(
    1 - \frac{g}{2}f_2\left(r R^2\right)
    + O\left(g^2\right)\right) + \\
  & \frac{8\pi^4}{81} g^3 m^6 R^2\left(
    f_3\left(r R^2\right)
    + O\left(g\right)\right) + O\left(g^4m^8\right) \Bigg)\,.
\end{split}
\end{equation}
Here the coefficients $r$ and $g$ are the renormalized counterparts to $r_0$
and $u_0$; the functions $f_1$, $f_2$ and $f_3$ are plotted in
fig.~\ref{fig:f_functions}, and an explicit expression (\ref{eq:f_functions})
is given below.

\begin{figure}
\begin{center}
\includegraphics[scale=0.75]{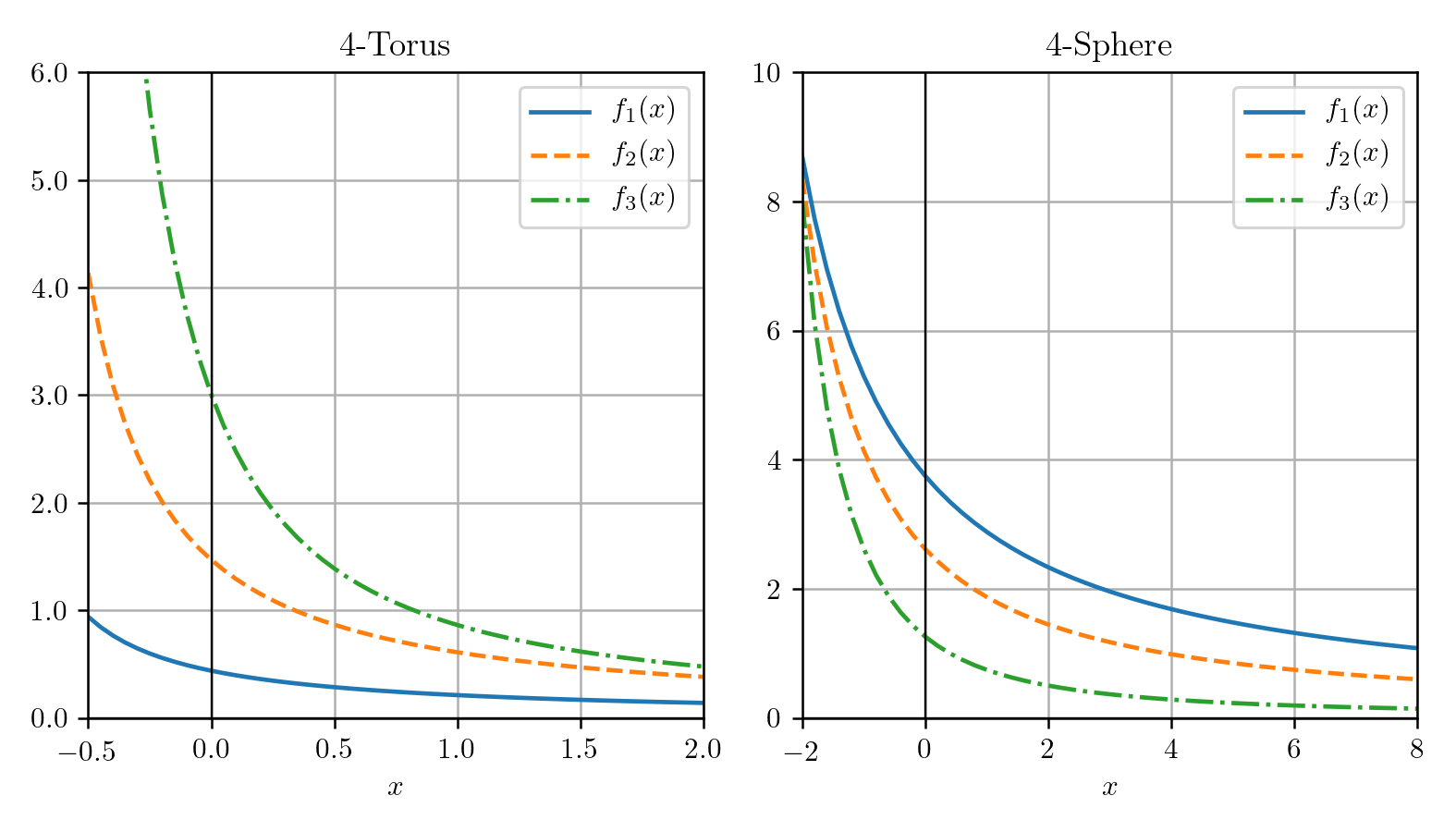}
\end{center}
\caption{\label{fig:f_functions} The functions $f_i$ that appear in
    (\ref{eq:log_pdf}). The functions go to zero as $x\to\infty$, and diverge
    for $x\to-1$ ($x\to-4$ for the sphere), signaling an instability of the
    perturbative vacuum.}
\end{figure}


The dimension-2 coupling $r$ controls the cross-over between the symmetric and
broken-symmetry phases. The expression (\ref{eq:log_pdf}) is valid for  $rR^2
\gtrsim -1/2$.  For lower values of $r$, the perturbative vacuum becomes
unstable, invalidating the perturbative expansion. For this reason, the
functions $f_i$ diverge as their argument approaches -1.  On the other hand,
$f_1$, $f_2$, $f_3$ all go to zero as their argument approaches positive
infinity, and hence for large $r$ the probability distribution of $m$ is a
zero-centered normal:
\begin{equation}
    p(m) \sim N e^{- \frac{1}{2} V r m^2}
  \quad \text{for} \quad
  rR^2 \gg 1\,.
\end{equation}

The dimensionless coupling $g$ is the parameter of the perturbative expansion.
The expansion is valid for $g \lesssim 1$, provided $rR^2$ is sufficiently far
from the bound discussed above.

In order to describe how $p(m)$ depends on $R$ at fixed bare couplings $r_0$,
$u_0$, it is necessary to account for renormalization effects. The specific
renormalization scheme we used is described in section \ref{sect:derivation},
and (\ref{eq:log_pdf_with_mu}) gives expression for $p(m)$ evaluated at a
generic renormalization scale $\mu$. However, for simplicity, we chose to
evaluate (\ref{eq:log_pdf}) at the scale $\mu^2 = r + R^{-2}$.  This choice of
$\mu$ is optimal for the reliability of perturbation theory, because it avoids
the emergence of large logarithms over the widest possible range of parameters.

At one loop, the Callan-Symanzik equations for $r$ and $g$ are:
\begin{equation}
  \label{eq:callan_symanzik}
  \mu \frac{\dd g}{\dd \mu} = g^2\,,\quad
  \mu \frac{\dd r}{\dd \mu} = \frac{1}{3}g r\,.
\end{equation}

These can be integrated and combined with the condition $\mu^2 = r + R^{-2}$ to
obtain a system of equations\footnote{There are of course many alternative,
arguably simpler, solutions that differ by sub-leading orders in an expansion in
$g(R_1)$. The one displayed here is the exact solution to
(\ref{eq:callan_symanzik}).} connecting the renormalized couplings at two
different values of $R$:
\begin{equation}
  \label{eq:renormalized_relation}
  \frac{g(R_1)}{g(R_2)} = \frac{r(R_1)^3}{r(R_2)^3} = 
  1 - \frac{1}{2} g(R_1)\log\frac{r(R_2) + R_2^{-2}}{r(R_1) + R_1^{-2}}\,.
\end{equation}

\begin{figure}
\begin{center}
\includegraphics[scale=0.75]{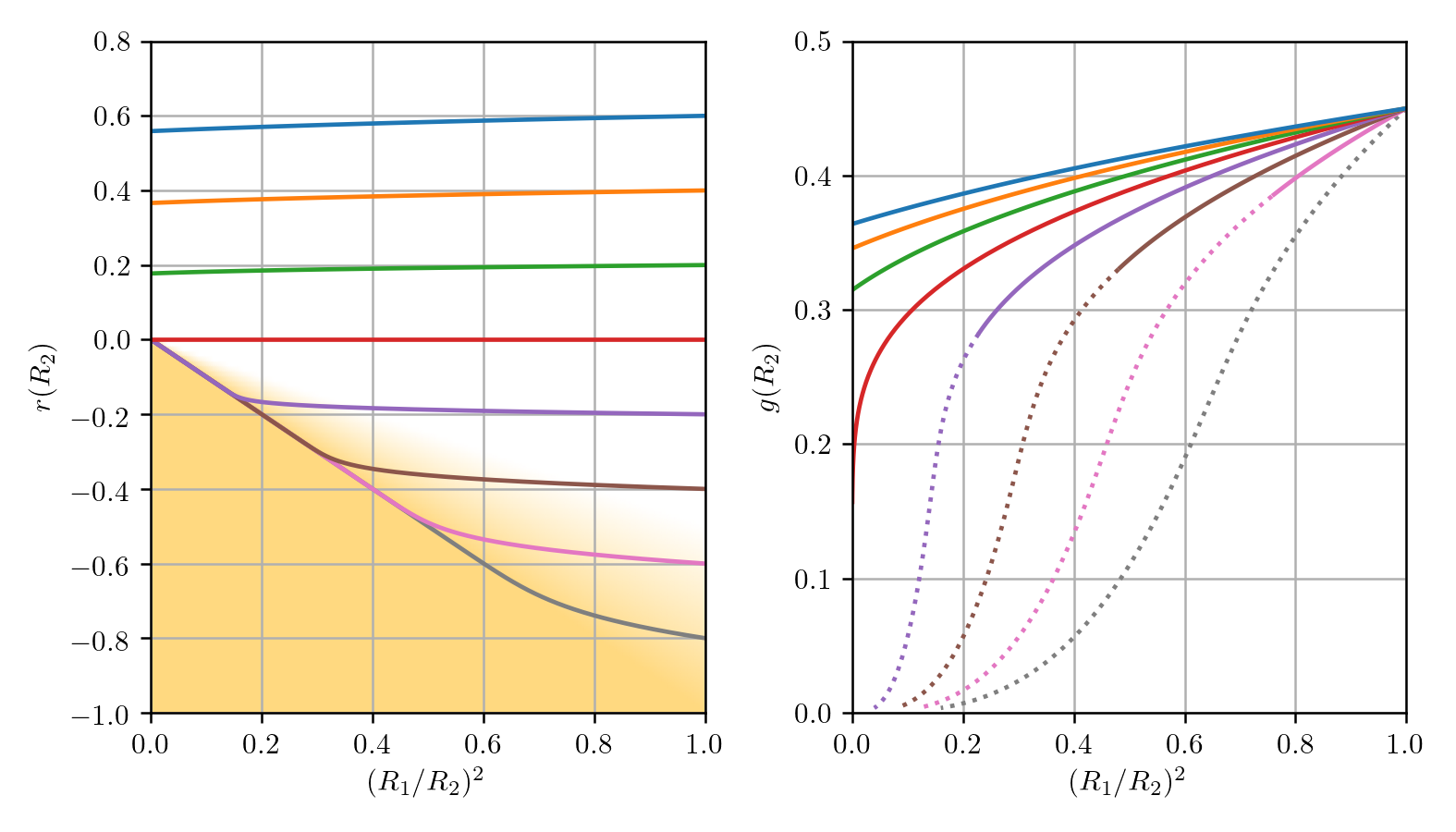}
\end{center}
    \caption{\label{fig:renormalization_group} A few solutions to
    (\ref{eq:renormalized_relation}) with $g(R_1) = 0.45$.  In the left plot,
    the shaded region shows where the perturbative vacuum becomes unstable,
    invalidating the perturbative expansion. Similarly, in the right plot, the
    lines become dotted outside of the perturbative region. Note how the sign
    of $r$ is preserved and $r = 0$ is a solution.}
\end{figure}

A few solutions to this system of equations are shown in
fig.~\ref{fig:renormalization_group}. Within the perturbative regime $g
\lesssim 1$, $r R^2 \gtrsim -1/2$, the coupling $r$ varies little with
$R$, and, as is clear from the differential form (\ref{eq:callan_symanzik}),
the sign of $r$ is always preserved. Thus we conclude that the critical point
is at $r = 0$, and the symmetric phase is realized for $r > 0$.

At the critical point, the renormalized coupling $g$ follows the simpler
Callan-Symanzik equation:
\begin{equation}
    \frac{1}{g(R_2)} - \frac{1}{g(R_1)} = \log\frac{R_2}{R_1}\,,
\end{equation}
and hence, as system size grows, the renormalized coupling $g$ goes to zero as
$1 / \log(R)$.  In this regime, the quartic term in (\ref{eq:log_pdf})
dominates all the others. This is perhaps most evident if the PDF is expressed
in terms of the rescaled quantity $\bar{m} = g^{\frac{1}{4}} m$, whose variance
remains finite as $g\to0$. Thus we conclude that, at the critical point, for
sufficiently large system size:
\begin{equation}
    \label{eq:log_pdf_critical_point}
    -\log p(m) \sim \mathrm{const} + \frac{2\pi^2}{9} V gm^4\,.
\end{equation}

In Monte Carlo simulations, the qualitative behavior of the distribution of the
magnetization is often characterized by measuring the so-called Binder
cumulant \cite{Binder1981}:
\begin{equation}
    \label{eq:binder_cumulant}
    U = 1 - \frac{\left\langle m^4 \right \rangle}
    {3\left\langle m^2 \right \rangle^2}\,.
\end{equation}

This quantity is constructed to be independent of the overall scale of $m$,
and to be zero if $m$ is normally distributed. From
(\ref{eq:log_pdf_critical_point}) we conclude that, on a 4-Torus, at the
critical point:
\begin{equation}
    \label{eq:critical_binder_cumulant}
    U = 1 - \frac{4\,\Gamma\left(\frac{5}{4}\right)^2}
    {3\,\Gamma\left(\frac{3}{4}\right)^2} 
    = 0.27052\ldots
\end{equation}

In fig.~\ref{fig:binder_cumulant}, we show a comparison of the Binder cumulant
computed from (\ref{eq:log_pdf}) near the critical point, and the results of
Monte Carlo simulation of the $4D$ Ising model. The agreement is excellent
except for the smallest system size $L = 8$. Notice how slowly the finite size
Binder cumulant approaches the asymptotic value
(\ref{eq:critical_binder_cumulant}).

\section{Derivation on the 4-torus}
\label{sect:derivation}

We now describe briefly how the result (\ref{eq:log_pdf}) is obtained. The
perturbative approach is similar to the computation of the effective
action, as in \textit{e.g.} \cite{WeinbergEffectiveAction}, except that we are
interested in the whole probability distribution of $m$, instead of
just the expected value. The main difficulty lies in evaluating the loop
integrals at finite size.

The average magnetization $m$ is proportional to the zero-momentum mode of the
field:
\begin{equation}
    \phi(x) = \frac{1}{V} \sum_{n \in \mathbb{Z}^4} \phi_{n} 
    e^{i \frac{n\cdot x}{R}}\,;\quad m = \frac{1}{V}\int\dd^4 x\ \phi(x) = \frac{1}{V}\phi_{n = 0}\,.
\end{equation}

Because of the delta function in (\ref{eq:pdf_definition}), the zero-mode
becomes an external field, whereas all other modes are still part of the
functional integral. Separating the zero-mode from the other modes in the
action yields:
\begin{equation}
\begin{split}
    S(m, \phi) =& 
    V \left(\frac{1}{2} r_0 m^2 + \frac{u_0}{24} m^4\right) +
    \frac{1}{2V} \sum_{n} \left(\frac{n^2}{R^2} + r_0 +
      \frac{1}{2} u_0m^2 \right)\phi_{n} \phi_{-n} +\\
    &\frac{u_0}{6V^2} m 
        \sum_{n_1, n_2} \phi_{n_1} \phi_{n_2} \phi_{-n_1 -n_2} +
    \frac{u_0}{24V^3} \sum_{n_1, n_2, n_3} \phi_{n_1} \phi_{n_2}
        \phi_{n_3} \phi_{-n_1 -n_2 -n_3}\,,
\end{split}
\end{equation}
where all summations now are over $\mathbb{Z}^4 \setminus \{0\}$.

We introduce renormalized couplings
\begin{equation}
    u_0 = u(1 + u\delta u)\,;\quad r_0 = r(1 + u \delta r)\,,
\end{equation}
and we obtain the following edges and vertices in the diagrammatic expansion:
\begin{displaymath}
\begin{tikzpicture}[scale=0.5, baseline=(current bounding box.center)]
    \draw[thick] (-1, 0) -- (1, 0);
    \node[left] at (-1, 0) {$n_1$};
    \node[right] at (1, 0) {$n_2$};
\end{tikzpicture}
    \quad
    \frac{V}{n_1^2R^{-2} + r}\delta_{n_1 + n_2}
\quad\quad
\begin{tikzpicture}[scale=0.5, baseline=(current bounding box.center)]
    \draw[densely dashed, thick] (-1, -1) -- (0, 0) -- (-1, 1);
    \draw[thick] (+1, -1) -- (0, 0) -- (+1, 1);
    \node[above right] at (1, 1) {$n_1$};
    \node[below right] at (1, -1) {$n_2$};
\end{tikzpicture}
    -\frac{um^2}{4V}\delta_{n_1 + n_2}
\end{displaymath}
\begin{displaymath}
\begin{tikzpicture}[scale=0.5, baseline=(current bounding box.center)]
    \draw[densely dashed, thick] (-1, -1) -- (0, 0);
    \draw[thick] (0, 0) -- (-1, 1);
    \draw[thick] (+1, -1) -- (0, 0) -- (+1, 1);
    \node[above left] at (-1, 1) {$n_1$};
    \node[above right] at (1, 1) {$n_2$};
    \node[below right] at (1, -1) {$n_3$};
\end{tikzpicture}
    -\frac{um}{6V^2}\delta_{n_1 + n_2 + n_3}
\quad\quad
\begin{tikzpicture}[scale=0.5, baseline=(current bounding box.center)]
    \draw[thick] (-1, -1) -- (1, 1);
    \draw[thick] (-1, 1) -- ( 1, -1);
    \node[above left] at (-1, 1) {$n_1$};
    \node[above right] at (1, 1) {$n_2$};
    \node[below right] at (1, -1) {$n_3$};
    \node[below left] at (-1, -1) {$n_4$};
\end{tikzpicture}
    -\frac{u}{24V^3}\delta_{n_1 + n_2 + n_3 + n_4}
\end{displaymath}
plus additional vertices associated with the counterterms $\delta r$, $\delta
u$ which we do not list for brevity.

The logarithm of the probability distribution of $m$ is the sum of all
connected diagrams:
\begin{displaymath}
\begin{split}
    \log p(m) = &\,
    \mathrm{const} -V \left(\frac{1}{2} r m^2 + 
    \frac{u}{24} m^4\right)\\
&+\ 
\begin{tikzpicture}[scale=0.5, baseline={([yshift=-.5ex]current bounding box.center)}]
    \draw[thick, densely dashed] (-1.75, -0.75) -- (-1, 0) -- (-1.75, 0.75);
    \draw[thick] (-1, 0) .. controls (-0.2, 0.8) and (0.5, 0.8) .. (0.5, 0) .. 
    controls (0.5, -0.8) and (-0.2, -0.8) .. (-1, 0);
\end{tikzpicture}
\ + \ 
\begin{tikzpicture}[scale=0.5, baseline={([yshift=-.5ex]current bounding box.center)}]
    \draw[thick, densely dashed] (-1.75, 0) -- (-0.75, 0);
    \draw[thick] (-0.75, 0) -- (0.75, 0);
    \draw[thick, densely dashed] (1.75, 0) -- (0.75, 0);
    \draw[thick] (0, 0) circle (0.75cm);
\end{tikzpicture}
\ + \ 
\begin{tikzpicture}[scale=0.5, baseline={([yshift=-.5ex]current bounding box.center)}]
    \draw[thick, densely dashed] (-2.75, -0.75) -- (-2, 0) -- (-2.75, 0.75);
    \draw[thick] (-2, 0) .. controls (-1.2, 0.8) and (-0.8, 0.8) .. (0, 0) .. 
    controls (-0.8, -0.8) and (-1.2, -0.8) .. (-2, 0);
    \draw[thick] (1.6, 0) .. controls (1.6, 0.8) and (0.8, 0.8) .. (0, 0) .. 
    controls (0.8, -0.8) and (1.6, -0.8) .. (1.6, 0);
\end{tikzpicture}
 \ + \ \cdots \\
&+\ 
\begin{tikzpicture}[scale=0.5, baseline={([yshift=-.5ex]current bounding box.center)}]
    \draw[thick, densely dashed] (-1.75, -0.75) -- (-1, 0) -- (-1.75, 0.75);
    \draw[thick, densely dashed] (1.75, -0.75) -- (1, 0) -- (1.75, 0.75);
    \draw[thick] (-1, 0) .. controls (-0.2, 0.8) and (0.2, 0.8) .. (1, 0) .. 
    controls (0.2, -0.8) and (-0.2, -0.8) .. (-1, 0);
\end{tikzpicture}
\ + \ 
\begin{tikzpicture}[scale=0.5, baseline={([yshift=-.5ex]current bounding box.center)}]
    \draw[thick, densely dashed] (-1.5, 0.75) -- (-0.75, 0) -- (-1.5, -0.75) ;
    \draw[thick] (0, 0) circle (0.75cm);
    \draw[thick, densely dashed] (0.53, 0.53) -- (1.25, 1.25);
    \draw[thick, densely dashed] (0.53, -0.53) -- (1.25, -1.25);
    \draw[thick] (0.53, -0.53) .. controls (0.2, -0.2) 
    and (0.2, 0.2) .. (0.53, 0.53);
\end{tikzpicture}
\ + \ 
\begin{tikzpicture}[scale=0.5, baseline={([yshift=-.5ex]current bounding box.center)}]
    \draw[thick, densely dashed] (-2.75, -0.75) -- (-2, 0) -- (-2.75, 0.75);
    \draw[thick, densely dashed] (2.75, -0.75) -- (2, 0) -- (2.75, 0.75);
    \draw[thick] (-2, 0) .. controls (-1.2, 0.8) and (-0.8, 0.8) .. (0, 0) .. 
    controls (-0.8, -0.8) and (-1.2, -0.8) .. (-2, 0);
    \draw[thick] (2, 0) .. controls (1.2, 0.8) and (0.8, 0.8) .. (0, 0) .. 
    controls (0.8, -0.8) and (1.2, -0.8) .. (2, 0);
\end{tikzpicture}
 \ + \ \cdots \\
& + \ 
\begin{tikzpicture}[scale=0.5, baseline={([yshift=-.5ex]current bounding box.center)}]
    \draw[thick, densely dashed] (-1.5, 0.75) -- (-0.75, 0) -- (-1.5, -0.75) ;
    \draw[thick] (0, 0) circle (0.75cm);
    \draw[thick, densely dashed] (1.40, -0.92) -- (0.375, -0.65) -- (0.1, -1.67);
    \draw[thick, densely dashed] (1.40, 0.92) -- (0.375, 0.65) -- (0.1, 1.67);
\end{tikzpicture}
 \ + \ \cdots \,,
\end{split}
\end{displaymath}
where again we left out all diagrams involving the counterterms for brevity. 

Retaining only one-loop diagrams, and employing a heat kernel regulator we have:
\begin{equation}
\begin{split}
    -\log p(m) = \mathrm{const} + V \Bigg( 
    &\frac{1}{2} m^2\left( r + u \left(\delta r + \frac{1}{2} I_1\right) + O(u^2)\right) + \\
    &\frac{u}{24} m^4 \left(1 + u \left(\delta u - \frac{3}{2} I_2\right) + O(u^2)\right) + \\
    &\frac{u^3}{48}m^6\left(I_3 + O(u)\right) + O(u^4 m^8) \Bigg)\,,
\end{split}
\end{equation}
where
\begin{equation}
    I_k = \frac{1}{V} \sum_{n\in\mathbb{Z}^4\setminus\{0\}} 
    \frac{e^{-s(n^2 R^{-2} + r)}}{\left(n^2 R^{-2} + r\right)^k} = 
    \frac{1}{(k - 1)!}\left(-\frac{\partial}{\partial r} - s\right)^{k-1} I_1
\end{equation}

The summation $I_1$ can be simplified substantially using the following trick:
\begin{equation}
\begin{split}
    I_1 &= \frac{1}{V} \int_{s}^{\infty} \dd t \sum_{n\in\mathbb{Z}^4\setminus\{0\}} 
    e^{-t(n^2 R^{-2} + r)}\\
    & = \frac{1}{V} \int_{s}^{\infty} \dd t\ e^{-t r}
        \left(\left(\sum_{m = -\infty}^{\infty} e^{-t m^2 R^{-2}}\right)^4 - 1\right) \\
    & = \frac{1}{V} \int_{s}^{\infty} \dd t\ e^{-t r}
        \left(\theta\!\left(t R^{-2}\right)^4 - 1\right)\,,
\end{split}
\end{equation}
where:
\begin{equation}
    \label{eq:theta}
    \theta(z) = \theta_3(0; e^{-z}) = \sum_{m = -\infty}^{\infty} e^{-m^2 z}\,.
\end{equation}
Here $\theta_3$ is a Jacobi theta function, but we will not need any of
its special properties beyond its asymptotic behavior, which can be easily
obtained from the definition.

We now extract the divergent and finite parts of $I_1$ as $s\to 0$ by
subtracting under the integral a function that has the same asymptotic behavior
as the integrand for $t \to 0$, but whose integral can be computed in closed
form. In order to do that, we need the asymptotic behavior of $\theta(z)$ for
$z\to0$, which can be obtained from its definition using the Euler-Maclaurin
formula:
\begin{equation}
    \theta(z) \sim \sqrt{\frac{\pi}{z}} + o(z^k) \quad\mathrm{for}\, z\to0\,.
\end{equation}

Thus we have:
\begin{equation}
\begin{split}
    I_1 &= \frac{1}{16\pi^2}\left(\int_{s}^{\infty}\dd t\, e^{-t \mu^2} t^{-2}\left(1 + t (\mu^2 - r)\right)
    +\frac{1}{R^2} f(r R^2, \mu R) + O(s)\right)\\
    &= \frac{1}{16\pi^2}\left(\frac{1}{s} + r\left(\log s\mu^2 + \gamma_{E}\right) - 
    \mu^2 +\frac{1}{R^2} f(r R^2, \mu R) + O(s)\right)\,,
\end{split}
\end{equation}
where $\mu$ is a renormalization scale that can be chosen at will, and:
\begin{equation}
    \label{eq:f_definition}
    f(\bar{r},\bar{\mu}) = \int_{0}^{\infty} \dd z\ e^{-z \bar{r}} \left(\frac{\theta(z)^4 - 1}{\pi^2} - 
    \frac{1 + z\left(\bar{\mu}^2 - \bar{r}\right)}{z^2}e^{-z\left(\bar{\mu}^2 - \bar{r}\right)}\right).
\end{equation}

We set the counterterms to:
\begin{align}
    &\delta r = -\frac{1}{32\pi^2}\left(\frac{1}{s} + r\left(\log s \mu^2 + \gamma_E - 1\right)\right)\,,\\
    &\delta u = -\frac{3}{32\pi^2}\left(\log s\mu^2 +\gamma_E + 1\right)\,.
\end{align}
The Callan-Symanzik equations (\ref{eq:callan_symanzik}) follow from this
subtraction choice.

Finally, we obtain:
\begin{equation}
\begin{split}
    \label{eq:log_pdf_with_mu}
    -\log p(m) =& \mathrm{const}+ V \Bigg(
    \frac{1}{2} m^2\left( r + \frac{u}{32\pi^2R^2}  \left(f(rR^2, \mu R) + (r - \mu^2)R^2\right) + O(u^2)\right) + \\
    &\frac{u}{24} m^4 \left(1 + \frac{3u}{32\pi^2} f^{(1, 0)}(rR^2, \mu R) + O(u^2)\right) + \\
    &\frac{u^3}{48}m^6\left(\frac{R^2}{32\pi^2}f^{(2, 0)}(rR^2, \mu R) + O(u)\right) + O(u^4 m^8) \Bigg)\,,
\end{split}
\end{equation}
from which (\ref{eq:log_pdf}) is obtained by setting $g = \frac{3 u}{16
\pi^2}$ and $\mu^2 = r + R^{-2}$. This last choice is motivated as follows. The
subtraction in (\ref{eq:f_definition}) is similar to:
\begin{equation}
    \int_0^{\infty} \dd z\, \frac{e^{-a z} - e^{-b z}}{z} = \log \frac{b}{a}\,.
\end{equation}
When the large-$z$ asymptotic behavior of the two terms is not well
matched, the integral becomes large in magnitude, making the perturbative
expansion less reliable. With the choice $\mu^2 = r + R^{-2}$, both terms in
(\ref{eq:f_definition}) have the same asymptotic behavior $\sim e^{-z(\bar{r} +
1)}$, thus avoiding the large log problem over the widest possible range of
parameters.

For completeness, let us display explicitly the functions $f_i$ that
parametrize (\ref{eq:log_pdf}):
\begin{align}
\label{eq:f_functions}
    &f_1(\bar r) = f(\bar r, \sqrt{\bar r + 1}) = 
    \int_{0}^{\infty} \dd z\ e^{-z \bar{r}} \left(\frac{\theta(z)^4 - 1}{\pi^2} - 
    \frac{(1 + z) e^{-z}}{z^2}\right)\,,\\
    &f_2(\bar r) = f^{(1, 0)}(\bar r, \sqrt{\bar r + 1}) = 
    \int_{0}^{\infty} \dd z\ e^{-z \bar{r}} z \left(\frac{\theta(z)^4 - 1}{\pi^2} - 
    \frac{e^{-z}}{z^2}\right)\,,\\
    &f_3(\bar r) = f^{(2, 0)}(\bar r, \sqrt{\bar r + 1}) = 
    \int_{0}^{\infty} \dd z\ e^{-z \bar{r}} z^2 \frac{\theta(z)^4 - 1}{\pi^2}\,,
\end{align}
and $\theta$ is given by (\ref{eq:theta}).

\section{Results for the 4-sphere}
\label{sec:four_sphere}
It is possible to obtain $p(m)$ on the 4 sphere as well, with
similar methods. Here we highlight the main differences from the 4 torus.

Curved manifolds allow for an additional renormalizable coupling: 
\begin{equation}
    \Delta S = \int_{\mathcal{M}} \frac{1}{2}\xi_0 \phi^2 \mathcal{R}\,,
\end{equation}
where $\mathcal{R}$ is the scalar curvature and $\xi_0$ is a dimensionless bare
coupling.  The free theory is Weyl invariant if $\xi_0 = \frac{1}{6}$. In the
presence of interactions, the coupling needs to be renormalized, and its
renormalized counterpart $\xi$ becomes a running coupling. At one loop, the
Callan-Symanzik equation for $\xi$ is:
\begin{equation}
    \mu \frac{\dd \xi}{\dd \mu} =  \frac{1}{3} g \left(\xi - \frac{1}{6}\right)\,.
\end{equation}

From this expression it seems that $\xi$ can be set to the critical value
$\frac{1}{6}$ at all energy scales. However, this turns out to be an illusion:
at higher orders in perturbation theory the Callan-Symanzik equation
becomes inhomogeneous \cite{Brown1980}. Therefore, on a curved manifold, $\xi$
is simply another free parameter of the scalar field theory.

The probability distribution of the total magnetization on the 4-sphere is
still given by (\ref{eq:log_pdf}), with the substitution:
\begin{equation}
    rR^2 \to \frac{\xi}{12} + r R^2\,,
\end{equation}
where $R$ is now the radius of the sphere, and with the functions $f_i$ defined
as:
\begin{align}
    &f_1(\bar{r}) = (2 - \bar{r})\int_{0}^{\infty} \dd z\ e^{-z(\bar{r} + 4)} H(z) + \frac{7}{3}\,,\\
    &f_2(\bar{r}) = \int_{0}^{\infty} \dd z\ e^{-z(\bar{r} + 4)} 
    \left(1 + z (2 - \bar{r})\right) H(z) + \frac{6}{\bar{r} + 4}\,,\\
    &f_3(\bar{r}) = \int_{0}^{\infty} \dd z\ e^{-z(\bar{r} + 4)} 
    \left(2 + z (2 - \bar{r})\right)z H(z) + \frac{\bar{r} + 10 }{(\bar{r} + 4)^2}\,,\\
\end{align}
where:
\begin{equation}
    H(z) = \sum_{\ell = 1}^{\infty} (2\ell + 3)e^{-z (\ell(\ell + 3) - 4)} - \frac{1}{z}\,.
\end{equation}

The functions $f_i$ for the sphere are also displayed in
fig.~\ref{fig:f_functions}. They diverge for $\bar{r} \to -4$, signaling the
instability of the perturbative vacuum, and they go to zero for $\bar{r} \to
\infty$.

\end{document}